\documentclass{easychair}

\usepackage{doc}

\usepackage{url}
\usepackage{hyperref}
\usepackage{listings}
\usepackage{lastpage}
\usepackage{fancyhdr}
\usepackage{latexsym}
\usepackage{geometry}

\usepackage{graphicx}
\usepackage[utf8]{inputenc}
\usepackage{subfigure}
\usepackage{multirow}
\usepackage{rotating}
\usepackage{pdflscape}

\title{A Blockchain-based Educational Record Repository%
}

\author{
Emanuel E. Bessa\inst{1}\thanks{Bessa, E. is with UNIFACS IPQoS research group}
\and
Joberto S. B. Martins\inst{2}\thanks{Prof. Dr. Martins, J. is with UNIFACS IPQoS and NUPERC research groups}
}

\institute{
  Salvador University,
  Salvador, Bahia, Brazil\\
  \email{emanuel.bessa@hotmail.com}
\and
   Salvador University,
  Salvador, Bahia, Brazil\\
   \email{joberto.martins@unifacs.br}
 }

\authorrunning{Bessa and Martins}

\titlerunning{7th International Workshop on ADVANCEs in ICT, January~2019}

\begin{document}

\maketitle

\begin{abstract}
 The Blockchain technology was initially adopted to implement various cryptocurrencies. Currently, Blockchain is foreseen as a general purpose technology with a huge potential in many areas. Blockchain-based applications have inherent characteristics like authenticity, immutability and consensus. Beyond that, records stored on Blockchain ledger can be accessed any time and from any location. Blockchain has a great potential for managing and maintaining educational records. This paper presents a Blockchain-based Educational Record Repository (BcER\textsuperscript{2}) that manages and distributes educational assets for academic and industry professionals. The BcER\textsuperscript{2} system allows educational records like e-diplomas and e-certificates to be securely and seamless transferred, shared and distributed by parties.
\end{abstract}

% The table of contents below is added for your convenience. Please do not use
% the table of contents if you are preparing your paper for publication in the
% EPiC Series or Kalpa Publications series

%TABLE OF CONTENTS
%\setcounter{tocdepth}{2}
%{\small
%\tableofcontents}

%\section{To mention}
%
%Processing in EasyChair - number of pages.
%
%Examples of how EasyChair processes papers. Caveats (replacement of EC
%class, errors).

%------------------------------------------------------------------------------

%\begin{IEEEkeywords}
%Blockchain, Security, Authenticity, Integrity, Distributed Ledger, Secure Transaction, Hash, Educational Record, Educational Repository, Educational Asset, Educational Record Distribution.
%\end{IEEEkeywords}

\section{Introduction}

Blockchain is a technology  considered by many to be something as relevant as the rise of the Internet. There have been experiments with blockchains since the early 1990’s, but it was only in 2008, with the release of a white paper by an individual or group of individuals under the pseudonym of Satoshi Nakamoto, that blockchains gained wide adoption \cite{grech_blockchain_2017}.

Educational records are used worldwide and, from the user point of view, is an important asset for individuals pledging for scholarships, jobs and professional and academic visibility in general. Currently, our educational records management systems are mostly physically localized, require specific and non-trivial procedures to access information, are in many cases unreliable and, finally, do not follow or have any educational standards.

With the blockchain capabilities and citizen's global visibility perspective in mind this paper presents a Blockchain-based Educational Records Repository (BcER\textsuperscript{2}). BcER\textsuperscript{2} is intended to allow any individual to be able to store educational records and access multiple type of educational records with authenticity on a worldwide basis. It is a system to ensure educational records distributed management and access with inherent more security like authenticity and privacy.

In the next part of this article, section 2 summarizes the fundamental aspects of Blockchain technology. Section 3 indicates the relevant work being done related to Blockchain and BcER\textsuperscript{2} system. Section 4 describes the architecture, entities, components and implementation of the BcER\textsuperscript{2} system. Section 5 presents a proof-of-concept of the BcER\textsuperscript{2} implementation. Finally, section 6 presents the final considerations and future work.

\section{Fundamental aspects of Blockchain Technology} \label{sec:fundamental}

A blockchain is essentially a distributed database of records that keeps potentially all kind of data, like transactions, contracts and events. All information handling takes place across a peer-to-peer network and is maintained chronologically in digital blocks. These basic features and capabilities make Blockchain transparent, secure, decentralized and with almost unlimited storage capacity\cite{crosby_blockchain_2016}.

Blockchain uses the concept of hashing. The "hash" is a block signature and considers all data and transactions involved. In summary, a cryptography hash function takes a input string and turns it into a unique n-digit string  \cite{grech_blockchain_2017}.

\subsection{Key Blockchain Characteristics}

With traditional methods for recording transactions and tracking assets, participants on a network keep their own ledgers and other records. This traditional method can be expensive, partially because it involves intermediaries that charge fees for their services. It’s clearly inefficient due to delays in executing agreements and the duplication of effort required to maintain numerous ledgers. It’s also vulnerable because if a central system (for example, a bank) is compromised due to fraud, cyberattack, or a simple mistake, the entire business network is affected. To solve or improve traditional method Blockchain has a set of key characteristics: consensus, provenance, immutability and finality \cite{grech_blockchain_2017} \cite{baru_blockchain:_2018}.

All relevant participants make decisions by consensus, in this process most participants must agree that a transaction is valid. This goal is achieved through the implementation of consensus algorithms. Each network enforces the conditions under which transactions are performed or the exchange of assets may occur. Provenance guarantees that participants know where the asset came from and how its ownership has changed over time. With immutability, no participant can tamper with a transaction after it has been recorded to the ledger. If a transaction is in error, a new transaction must be used to reverse the error, and both transactions are then visible. With finality, a single shared ledger provides one place to go to determine the ownership of an asset or the completion of a transaction.

%\begin{itemize}
%\item Consensus:  All relevant participants make decisions by consensus, in this process most participants must agree that a transaction is valid. This goal is achieved through the implementation of consensus algorithms. Each network enforces the conditions under which transactions are performed or the exchange of assets may occur. 

%\item Provenance: Participants know where the asset came from and how its ownership has changed over time. 

%\item Immutability: No participant can tamper with a transaction after it has been recorded to the ledger. If a transaction is in error, a new transaction must be used to reverse the error, and both transactions are then visible.

%\item Finality: A single shared ledger provides one place to go to determine the ownership of an asset or the completion of a transaction.
%\end{itemize}

\subsection{Why "Blockchain" to Manage and Improve Visibility for Educational Records} \label{Problema}

An “education record” in the context of this paper is a record containing files, documents, and other materials which  \cite{grech_blockchain_2017}: i) Contains information directly related to the academic historical of a student or a professional; and ii) From a local perspective are typically maintained by an educational institution or by other entity acting for such institution.

%\begin{itemize}
%    \item Contains information directly related to the academic historical of a student or a professional; and
%    \item From a local perspective are typically maintained by an educational institution or by other entity acting for such institution.
%\end{itemize}

There are significant advantages and benefits in using a Blockchain-based educational repository \cite{blockchain_blockchain_2018} \cite{sheeraz_blockchain_2018}: i) Educational records (e-diplomas, e-certificates, other) uploaded and managed on the Blockchain ledger are more secure and resistant to “physical wear and tear” than paper documents \cite{blockchain_blockchain_2018}; ii) Educational records are seamless and efficiently transferred and shared among parties (universities, schools and employers) fostering worldwide visibility; and iii) Educational records stored on the blockchain can be accessed any time, from any location.

%\begin{itemize}
%    \item Educational records (e-diplomas, e-certificates, other) uploaded and managed on the Blockchain ledger are more secure and resistant to “physical wear and tear” than paper documents \cite{blockchain_blockchain_2018};
%    \item Educational records are seamless and efficiently transferred and shared among parties (universities, schools and employers) fostering worldwide visibility; and
%    \item Educational records stored on the blockchain can be accessed any time, from any location.
%\end{itemize}

In summary, educational records managed by Blockchain technology stimulate the knowledge/reward principle, makes credentials more trustworthy and keeps educational records safe and easy to access \cite{blockchain_blockchain_2018}.

\section{Related Work}\label{sec:relatedwork}

In recent years, blockchain technology has been widely used as the basic construct for crypto-coins such as Bitcoin \cite{nakamoto_bitcoin_2008}.

MIT has a system for building Blockchain-based applications that issues and verifies official records called "Blockcerts Wallet". It allows, for instance, the creation of a certificate wallet for students to receive  virtual diplomas via their smart devices  \cite{lab_what_2016}. Different from Blockchain-based Educational Records Repository (BcER\textsuperscript{2}), MIT Blockcerts Wallet system is a building application platform that has a similar target in terms of allowing educational records creation and dissemination using Blockchain.

New promisingly Blockchain-based solutions include 'intelligent contracts'. Ethereum, discussed in \cite{reiff_blockchain_2018}, allows the creation of contracts that are self-managed.  Contracts are triggered by an event such as passing an expiration date or achieving a specific price goal. In response, the smart contract manages itself by making adjustments as needed and without the input of external entities \cite{azam_what_2018}.

\section{The Blockchain-based Educational Records Repository (BcER\textsuperscript{2}) - Architecture and Components} \label{sec:architecture}

In order to meet the different alternatives of use, blockchain-based applications can be implemented using 3 types of general structures as follows: i) Public Blockchain; ii) Private Blockchain; and iii) Consortium Blockchain.

%\begin{itemize}

%\item Public Blockchain; 
%\item Private Blockchain; and
%\item Consortium Blockchain. 
%\end{itemize}

The BcER\textsuperscript{2} repository adopted the consortium system since only authorized persons are able to create certificates records on the network. On the other hand, anyone can verify their authenticity. Thus, when registering an educational record, for example, the responsible for creating the record writes in the registry or in the database using its own private key. Users who want to check the veracity of the record must have a corresponding identifier number to be inserted into the system.

The Blockchain consortium is a semi-private and partially decentralized chain system, in this scenario the nodes are responsible for the validation of the transactions and how this happens depends on the implementation of the consensus methods. The form of access and consultation of the records can be public or private and the owner of the network is responsible for its configuration.

The BcER\textsuperscript{2} effective structure uses the basic steps and operation flow of a blockchain-based application: i) A transaction is requested by someone who has prior authorization and needs to create an educational record; ii) The request record transaction is sent to the nodes belonging to the BcER\textsuperscript{2} system; iii) The educational record transaction is verified by the ledger; and iv) A new block of data corresponding to the educational record transaction is accessed or created and annexed to the ledger becoming permanent and immutable completing the transaction.

%\begin{itemize}
%    \item A transaction is requested by someone who has prior authorization and needs to create an educational record;
%    \item The request record transaction is sent to the nodes belonging to the BcER\textsuperscript{2} system;
%    \item The educational record transaction is verified by the ledger; and
%    \item A new block of data corresponding to the educational record transaction is accessed or created and annexed to the ledger becoming permanent and immutable completing the transaction.
%\end{itemize}

\subsection{Blockchain-based Educational Records Repository (BcER\textsuperscript{2}) Entities}

The entities belonging to the BcER\textsuperscript{2} educational repository are the following: i) Assets; ii) Registers; iii) Transactions; and iv) Participants.

%\begin{itemize}
%    \item Assets;
%   \item Registers;
%    \item Transactions; and
%    \item Participants.
%\end{itemize}

%operation, data flow and steps is necessary to understand the entities involved in the proposed Blockchain.

An "asset" can be anything of value that will be kept securely by the educational repository. Educational records like certificates, diplomas, educational records an similar documents are BcER\textsuperscript{2} assets.

"Participants" are the educational organization representatives, students and people in general that are somehow interested in either distributing or accessing educational records. Participants are defined in the "business network model" adopted by the blockchain application process. For BcER\textsuperscript{2}, coordinators, students, and anyone else interested in having access to the educational records are the participants belonging to the network.  Each one has their specifically assigned functions, responsibilities and access restrictions.

"Transactions" are submitted by participants to create or access the assets held in the blockchain-based asset registries on the blockchain ledger. Transactions, in general, do belong to a business network and, as such, do require a "business network model". The business network model, from the blockchain system perspective,  define the operation involved with the assets.

"Registers" can be defined as the set of data that involve the assets, transactions and the participants. this set is what will be included in the block after the validation. Registers are new information that is added to the blockchain ledger.

In BcER\textsuperscript{2} educational records repository the addition of a "register" is carried out through the execution of steps. In these steps, the business model adopts the following "business" premises: i) The course coordinator or the educational institution representative are the authority to create new assets; and ii) Students and general public are participants that access the validated and secure educational assets maintained by the system.

%\begin{itemize}
%    \item The course coordinator or the educational institution representative are the authority to create new assets; and
%    \item Students and general public are participants that access the validated and secure educational assets maintained by the system.
%\end{itemize}

\begin{figure}
    \centering
    \includegraphics[width=.65\linewidth]{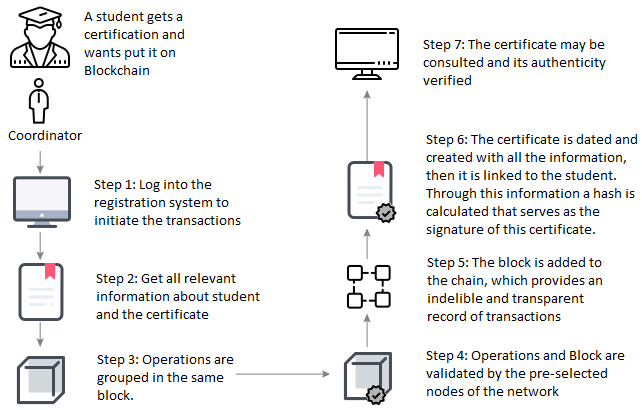}
    \caption{"Certificate Register" creation in BcER\textsuperscript{2}}
    \label{}
\end{figure}

The creation of a register is executed as illustrated in Figure 1: i) A coordinator proceeds to write a record to the Blockchain account, which means create the certificate with an identifier. In this process the coordinator selects the certificate and through its identifier number it is possible to link it to a student; ii) The record is saved and time-stamped in a block using arithmetic operations; iii) The block is subsequently validated by network pre-selected nodes through cryptography techniques; and iv) The block is dated and added to the block chain, so that all users can have access to the same chain since each node builds its own exemplary independently.

Once these steps have been executed, we can access educational records with authenticity and integrity by simply using a credential (ID Card) through a web browser.

\subsection{Blockchain-based Educational Records Repository (BcER\textsuperscript{2}) Business Network}

The business network is a fundamental definition for the BcER\textsuperscript{2} educational registry repository deployment.

In summary, it models the BcER\textsuperscript{2} "educational model", defining the existing assets, transactions and participants related to them. The business network defines the transactions that interact with assets. The model also includes the definition of participants who interact with assets and associates a unique identity, across multiple business networks. As described before, BcER\textsuperscript{2} is composed of assets, participants and transactions, with each of these entities modeled in relation to the educational operation.

\subsection{Blockchain-based Educational Records Repository (BcER\textsuperscript{2}) Components}

The basic components belonging to the the BcER\textsuperscript{2} educational records repository are illustrated in Figure 2 and basically reflect the business network adopted which is suitable for an educational records repository that registers, manages and provide access to them. 

\begin{figure}[ht]
    \centering
    \includegraphics[width=.7\linewidth]{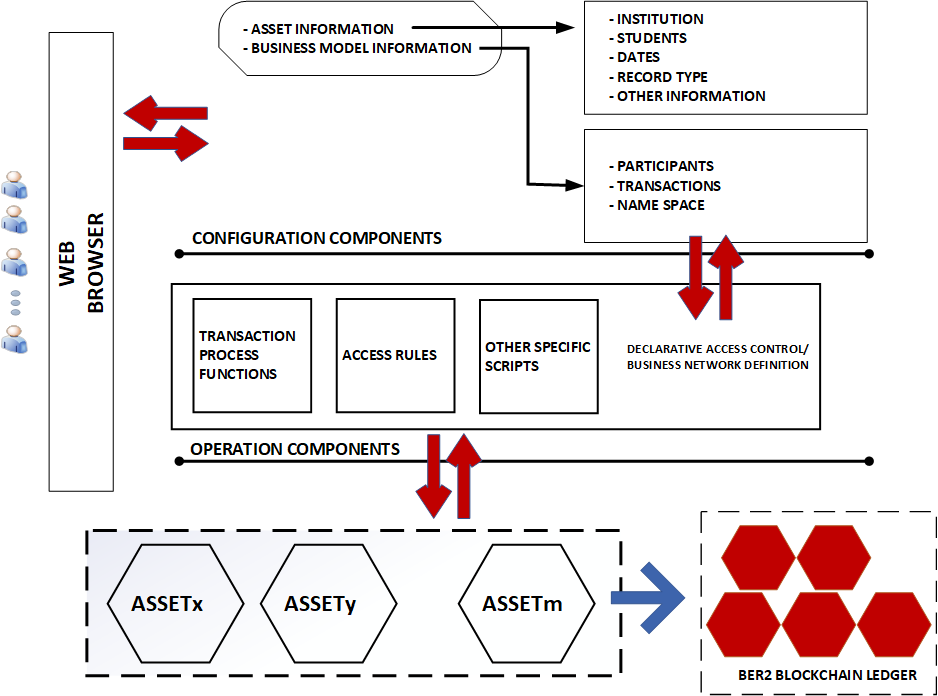}
    \caption{BcER\textsuperscript{2} basic components}
    \label{}
\end{figure}

The "asset information" component contains information related to the educational record being managed by BcER\textsuperscript{2}. This component is responsible for asset's definition and consistency. 

The "Business Model Information" component contains information related to the process involved in the asset management. It defines basically the participants, name space and transactions involved in the process.

The "Transaction Process Function" component contains the information concerning the specif functions invoke in the business model to manage the asset.

The "Access Rules" component contains, as the name suggests, the access rules including all priorities among participants involved in the business model adopted.

BcER\textsuperscript{2} managers and general users access the repository through a web browser using  identification cards (ID Cards) which includes connection profiles and credentials.

Assets are effectively deployed in the BcER\textsuperscript{2} blockchain ledger in this specif system by  the Hyperledger Composer framework described in section 4.4.

\subsection{Blockchain-based Educational Records Repository (BcER\textsuperscript{2}) Implementation} \label{sec:hyperledger}

The Hyperledger Composer \cite{Hyperledger_Composer_Documentation} was used to implement the BcER\textsuperscript{2} educational repository. Hyperledger Composer is an open source development tool set and framework aiming to support the development of blockchain applications. It allows the modeling of the business network and integrates existing systems components and data deploying as such the blockchain application.

The use case adopted by the actual implementation is initially intended to verify the authenticity of student's certificates generated by Salvador University (UNIFACS) under-graduation courses.

BcER\textsuperscript{2} is composed of assets, participants and transactions, with each of these entities being represented within Hyperledger framework as configuration files.

The .CTO hyperledger component is responsible for implementing the assets, participants, and transactions, including all relevant information. A Hyperledger Composer CTO file is composed of the following elements: i) A name-space with resources declaration; ii) Resources definition including assets, transactions, participants and events; and iii) Optional resource import declarations from other name-spaces.

%\begin{itemize}
%    \item A name-space with resources declaration;
%    \item Resources definition including assets, transactions, participants and events; and
%    \item Optional resource import declarations from other name-spaces.
%\end{itemize}

The .ACL hyperledger component provides declarative access control for the elements in the domain model. By defining access and control (ACL) rules you can determine which users/roles are permitted to create, read, update or delete elements in a business network's domain model.

The 'Business Network' definition, from the Hyperledger Composer perspective,  is composed by a set of model files defining assets, participants and transactions. The ".JS" script file is responsible for maintain a set of scripts. The scripts contain transaction process functions that implement the transactions defined in the 'Business Model'. Transaction processing functions are automatically invoked at run-time when transactions are submitted and their structure are composed by a JavaScript function.

\section{BcER\textsuperscript{2} Repository - Proof-of-Concept}
\label{sec:experimentacao}

The main objective of the BcER\textsuperscript{2} repository proof-of-concept is to validate the deployment of the business network and verify the operation steps of the repository by: i) Creating assets; and ii) Accessing them and verifying the effectiveness of credentials and other distribution and security aspects.

%\begin{itemize}
%    \item Creating assets; and
%    \item Accessing them and verifying the effectiveness of credentials and other distribution and security aspects.
%\end{itemize}

The proof-of-concept of the BcER\textsuperscript{2} repository operation was implemented by emulating participants as follows: i) 'Users' are the general public accessing educational records; and ii) The 'Coordinator' (Register Authority) is a UNIFACS authority creating educational record entries.

%\begin{itemize}
%    \item 'Users' are the general public accessing educational records; and
%    \item The 'Coordinator' (Register Authority) is a UNIFACS authority creating educational record entries.
%\end{itemize}

The experiment was executed using the following infrastructure: i) The BcER\textsuperscript{2} system runs on Notebook Core i7, 2.0 Ghz, 8 GB RAM, Ubuntu  Server  Operating  System 16.04  x64; and ii) The coordinator and users access the BcER\textsuperscript{2} system using any browser.

%\begin{itemize}
%    \item The BcER\textsuperscript{2} system runs on Notebook Core i7, 2.0 Ghz, 8 GB RAM, Ubuntu  Server  Operating  System 16.04  x64; and
%    \item The coordinator and users access the BcER\textsuperscript{2} system using any browser.
%\end{itemize}

The software components installed to run the BcER\textsuperscript{2} system are i) Node Version 8.12; ii) NPM Version 6.4.1; iii) Visual Studio Code Version Version 1.28; iv) Docker Engine Version 18.06; and v) Docker Composer Version 1.23.

%\begin{itemize}
%   \item Node Version 8.12
%                    \item NPM Version 6.4.1                      
%                    \item Visual Studio Code Version Version 1.28
%                    \item Docker Engine Version 18.06
%                    \item Docker Composer Version 1.23
%\end{itemize}

The proof-of-concept method and parameters used to validate the operation of BcER\textsuperscript{2} was the creation of a set of educational records followed by authenticity verification and access by the system administrator and distributed users. The experimental setup included the creation of 10 different educational records with course certificates and various nodes (N$\textgreater$10) simulating different users acting on the validation process (transactions) and verifying their authenticity. The results allowed the distributed access to educational records enabling verification through blockchain technology. Security access was also validated by trying to access educational records without the adequate credential. The scalability of the solution was not evaluated and will be addressed by future work.

\section{Final Considerations and Future Work}
\label{sec:conclusao}

We propose in this work to use Blockchain technology as a tool to provide a secure and efficient way to access certificate with authenticity. We argue that the proposed Blockchain BcER\textsuperscript{2} repository has the potential to support the education sector by providing better support for certificate management and distribution.

In the current scenario, the proposed application covers only the UNIFACS network for the purpose of managing diplomas and certificates issued by the university. The BcER\textsuperscript{2} application has the potential to cover additional areas in which digital certificates provide interesting opportunities such as: i) Corporate Training - Many large companies offer a multitude of training opportunities to their employees, but lack the systems to track and store results reliably. Current human resources systems often do not interact with corporate databases and there are no consistent standards for comparing skills and accomplishments; and ii) Workforce Development - There are millions of records and learning certificates, but there are no systems to manage them.

%\begin{itemize}
%\item Corporate Training - Many large companies offer a multitude of training opportunities to their employees, but lack the systems to track and store results reliably. Current human resources systems often do not interact with corporate databases and there are no consistent standards for comparing skills and accomplishments.
%\item Workforce Development - There are millions of records and learning certificates, but there are no systems to manage them. This is especially a problem for people with low qualifications, who often do not have recognized diplomas or degrees.
%\end{itemize}

In terms of future work, it is intended to evaluate the scalability issues and impacts associated with the deployment of a huge repository. Another aspect to be considered is to bring together stakeholders such as employers, students, teachers and contractors in a way that they interact with each other enabling wide-spread use of trustable e-certificates. A final target will be to adopt a fully standardized asset representation as an additional step towards a secure and decentralized way of conferring a wide-spread use of the system.

\section{Acknowledgements}
Authors thanks FAPESB (Fundação de Apoio à Pesquisa do Estado da Bahia) by the scientific initiation (IC) scholarship support.

%\printbibliography[maxnames=99]

\bibliographystyle{plain}
\bibliography{allbib.bib}

%\input{ADVANCE-Blockchain.bbl}

%\vskip -2\baselineskip plus -1fil

%\bibliographystyle{ieetr}
%\bibliography{allbib}

\includegraphics[width=.9in,height=1in,clip,keepaspectratio]{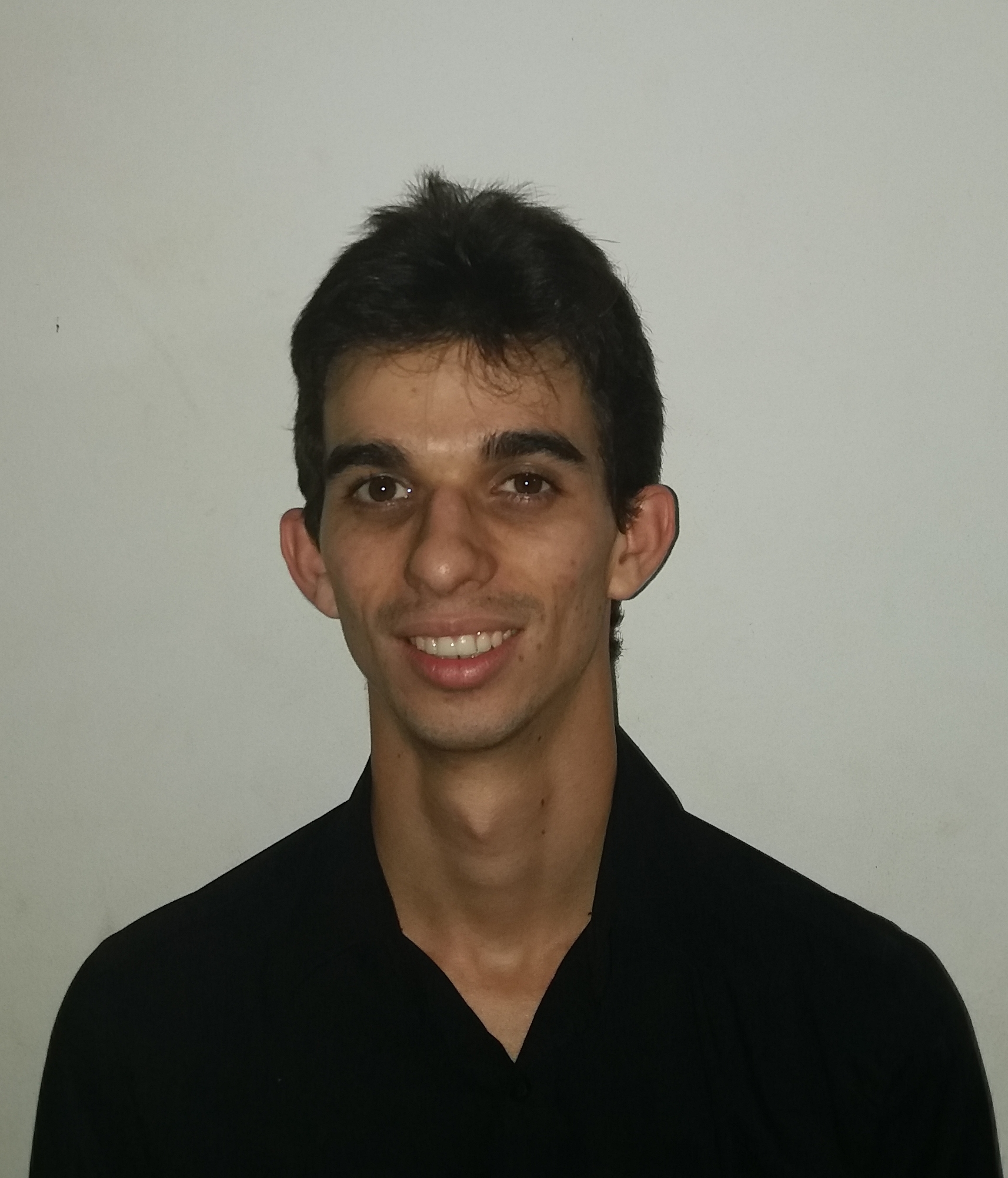}]
Emanuel E. Bessa -
Computer science student at Salvador University (UNIFACS). Research interest include Blockchain technology and Smart Cities. Emanuel is a member of IPQoS Research Group and RePAF Project (Resource Allocation Framework for MPLS, Elastic Optical Networks (EON), Network Function Virtualization (NFV) and IoT) at UNIFACS.

\hfill \break

\includegraphics[width=.9in,height=1in,,clip,keepaspectratio]{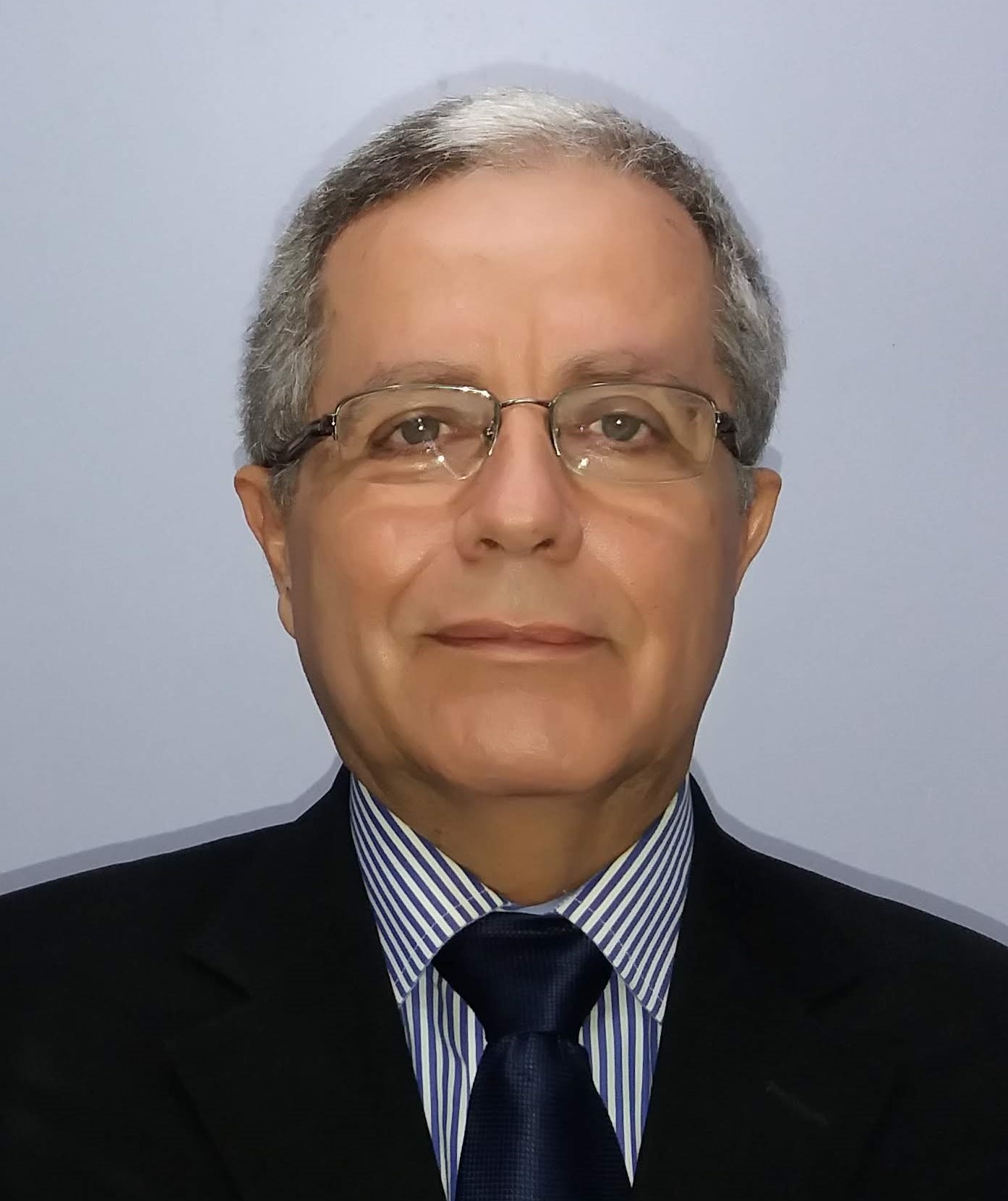}]
Prof. Dr. Joberto S. B. Martins -
PhD in Computer Science at Université Pierre et Marie Curie – UPMC, Paris (1986), PosDoc at ICSI/ Berkeley University (1995) and PosDoc Senior Researcher at Paris Saclay University – France (2016). Invited Professor at HTW – Hochschule für  Techknik und Wirtschaft des Saarlandes (Germany) (since 2004) and Université d'Evry (France). Full Professor at Salvador University on Computer Science, Head of NUPERC and IPQoS research groups with research interests on Cognitive Management, Artificial Intelligence, Resource Allocation, SDN/ OpenFlow, Internet of Things, Smart Grid and Smart Cities. Senior member of IEEE Smart Grid and IEEE Smart City Research Committees and former Chair of the IEEE CGAA (Committee on Global Accreditation Activities).

\end{document}